# Magnetic behavior of EuCu$_2$As$_2$: Delicate balance between antiferromagnetic and ferromagnetic order

**Kausik Sengupta, P.L. Paulose and E.V. Sampathkumaran**
*Tata Institute of Fundamental Research, Homi Bhabha Road, Colaba, Mumbai – 400005, India*
**Th. Doert and J.P.F. Jemetio**
*Institut für Anorganische Chemie, Technische Universität Dresden, Helmholtzstrasse 10, D-01062 Dresden, Germany*

The Eu-based compound, EuCu$_2$As$_2$, crystallizing in the ThCr$_2$Si$_2$-type tetragonal structure, has been synthesized and its magnetic behavior has been investigated by magnetization (M), heat-capacity (C) and electrical resistivity ($\rho$) measurements as a function of temperature (T) and magnetic field (H) as well as by $^{151}$Eu Mössbauer measurements. The results reveal that Eu is divalent ordering antiferromagnetically below 15 K in the absence of magnetic field, apparently with the formation of magnetic Brillouin-zone boundary gaps. A fascinating observation is made in a narrow temperature range before antiferromagnetism sets in: That is, there is a remarkable upturn just below 20 K in the plot of magnetic susceptibility versus T even at low fields, as though the compound actually tends to order ferromagnetically. There are corresponding anomalies in the magnetocaloric effect data as well. In addition, a small application of magnetic field (around 1 kOe at 1.8 K) in the antiferromagnetic state causes spin-reorientation effect. These results suggest that there is a close balance between antiferromagnetism and ferromagnetism in this compound.

## I. INTRODUCTION

Though a large number of ternary rare-earth compounds crystallizing in the ThCr$_2$Si$_2$-type (space group I4/mmm) have been known [1], the studies based on As compounds, particularly containing Eu, are scarce in the literature. This is presumably due to unfavorable intrinsic properties of Eu (instability in air) and As (high volatility and toxicity), which makes synthesis of such compounds by conventional melting techniques very difficult. In this article, we report the synthesis of one such compound, viz., EuCu$_2$As$_2$, by solid state reaction route and report its magnetic behavior by a variety of experimental methods. Incidentally, the synthesis of this compound is an offshoot of our unsuccessful attempts to prepare single-phase EuCuAs$_2$ (as a continuation of our studies [2] of compounds of the type RCuAs$_2$ (R= Rare-earths)). In 1995, Dűnner et al attempted the synthesis of this compound by direct reaction of the elements [3]. As no single crystals could be obtained by this method, a different synthetic route was adopted ((starting from Eu, CuCl, and As in molar ratios 15:14:16), which however resulted in crystals with a non-stoichiometry, viz., EuCu$_{1.75}$As$_2$; this material was found to order ferromagnetically at 32 K.

## II. EXPERIMENTAL DETAILS

Polycrystalline sample of EuCu$_2$As$_2$ was prepared by direct reaction starting from the elements, Eu (powder, 99.999%), Cu (powder, 99.9%), and As (sponge, 99.9%)) in the molar ratio of 1:2:2. The mixtures were filled in silica tubes, sealed under vacuum and placed into a box furnace. The samples were heated to 500 C with 8C/min, and subsequently annealed for 10-12 days at 900C with an intermediate grinding and cooled to room temperature thereafter. While initial handling (till sealing in quartz tube) was done in an argon box, the specimen obtained after a heat-treatment is found to be stable in air. Powder X-ray diffraction pattern (Cu K$_\alpha$) revealed that all the diffraction lines could be indexed to ThCr$_2$Si$_2$-type tetragonal structure without any evidence for any other extra phase. The lattice constants are found to be

a= 4.260 (1) Å and c = 10.203(1) Å. These values are slightly larger than those reported [1] for the non-stoichiometric composition, EuCu$_{1.75}$As$_2$ (a= 4.215 (1) Å and c = 10.185(1) Å). $^{151}$Eu Mössbauer measurements were carried at selected temperatures employing $^{151}$SmF$_3$ (E$_\gamma$ = 21.6 keV) source. Dc magnetic susceptibility ($\chi$) measurements were carried out in the temperature interval 1.8 to 300 K in the presence of few magnetic fields (H= 5 kOe as well as 100 Oe) and isothermal magnetization (M) behavior were also tracked at few fixed temperatures, employing commercial magnetometers. Temperature dependent heat-capacity (C) and electrical resistivity ($\rho$) measurements were performed in the absence as well as in the presence of magnetic fields employing a commercial Physical Property Measurements System (Quantum Design).

## III. RESULTS AND DISCUSSION
### A. Mössbauer spectroscopy

It is a well-known fact that, as far as Eu is concerned, Mössbauer spectroscopy is the ideal tool to get information about the valence state and magnetic ordering of Eu ions, as the isomer shift (IS) for divalent and trivalent Eu ions falls in a widely different range (about -8 to -14mm/s and 0 to +4 mm/s respectively with respect to EuF$_3$) and the spectrum undergoes hyperfine splitting with well-resolved features below the magnetic ordering temperature. Therefore the $^{151}$Eu Mössbauer spectra obtained at various temperatures 25, 15, 12 and 4.2 K are shown in figure 1. At 25 K that there is only one resonance line with an isomer shift of -11.5 mm/s, and at 15 K, there is a relatively marginal relaxation-broadening of the spectrum due to the onset of magnetic ordering. Further lowering of temperature to 12 K results in a splitting of the spectra and the spacing between the hyperfine-split lines increases further at 4.2 K. The isomer shift obtained by a conventional fitting of the spectra is found to be the same as that at 25 K. There is no significant resonance in the region around 0 mm/s and the trivalent Eu contribution to resonance, if present, is estimated to be not more than few percent (which usually appears if the alloy specimens are powdered in air for Mössbauer studies). These results establish that the Eu ions are essentially divalent and that these ions undergo magnetic ordering around 15 K. Due to difficulties in maintaining the temperature over several hours required for collecting the data, the uncertainty in the measured temperature is close 2 K. The magnitude of the internal hyperfine-field derived from the spectra at 4.2 K is found to be close to 320 kOe.

### B. Magnetic susceptibility and isothermal magnetization

The plot of inverse $\chi$(T) (Fig. 2, top) is found to be linear above 25 K and the value of the effective moment is found to be 7.9 $\mu_B$, which is very close to that expected for divalent Eu ions. The value of the paramagnetic Curie temperature ($\theta_p$) is 18.3 K. The positive sign of $\theta_p$ implies that the ferromagnetic exchange interaction is present. There is a sharp increase in $\chi$ as the temperature is lowered below 20 K, nearly at the same temperature as $\theta_p$, as though the compound tends to undergo ferromagnetic ordering. However, as the T is lowered below 17 K, a peak in $\chi$ appears at 15 K, that is, within a narrow range of T, with a significant fall for a further lowering of temperature. These observations imply that the magnetism of this compound could be quite complex.

To address the nature of the ground state magnetic structure, the isothermal M at several temperatures has been measured (see figure 3a for typical M(H) behavior). At all temperatures, M increases for initial applications of H (about 10 kOe), but there is a distinct evidence for a spin reorientation effect around 1 kOe at 1.8 and 10 K (see figure 3, inset), which can be explained only if the magnetic structure is of an antiferromagnetic type (in zero H) at these temperatures. Fig. 3b shows hysteresis loops at two temperatures, one (17 K) just above the $\chi$–peak temperature after the onset of magnetic ordering and the other at a lower temperature (5 K). Absence of hysteresis in M(H) curve at 5 K (that is, negligible coercivity within sensitivity limit of the magnetometer employed, of the order of few

Oe) is consistent with the low temperature antiferromagnetic ground state. It is also obvious from figure 3b that the shape of the hysteresis loop at 17 K looks different from that at 5 K in the sense that spin reorientation effect is suppressed and the response of M to initial application of H is much larger compared to that at 5 K – an observation supporting ferromagnetic correlations at 17 K. With respect to the high field behavior, say beyond 20 kOe, M at 1.8 K saturates and the saturation moment is close to 7 $\mu_B$, typical of ferromagnetically-coupled Eu ions. Thus, at high fields, this compound is ferromagnetic at low temperatures. The above conclusions can also drawn by looking at the behavior of Arrott plots ($M^2$ vs H/M, see figure 3c). At 1.8 K, $M^2$ is nearly constant for large values of H/M and the intercept on y-axis extrapolated from high fields remains positive for T ≤17 K, consistent with the fact that the high-field state is ferromagnetic. This intercept crosses sign between 17 and 20 K, that is, at the onset of magnetic ordering. A careful look at this figure at low fields is quite revealing. Initial slope of the plot at 14.3 K is almost infinite and, at lower temperatures, there is a curvature (towards higher values on the x-axis) before this 'infinite slope' appears. These features are typical of field-induced ferromagnets (see, for instance, Ref. 4), with 14.3 K being the upper limit in figure 3c for such a transition. These results therefore support our conclusion above that, below 14.3 K, the ground state of the compound (in zero magnetic field) is not ferromagnetic. In any case, the results discussed above establish that there is a close balance between antiferromagnetism and ferromagnetism in this compound. However, there is no evidence for spin-glass behavior, as we do not find any difference between zero-field-cooled (ZFC) and field-cooooled (FC) low-field $\chi(T)$ curves obtained in a field of 100 Oe (see figure 2, bottom). This was further confirmed by frequency independent peak in ac $\chi(T)$ plots, the shape of which qualitatively mimics those of dc low-field dc $\chi(T)$ (and hence not shown here).

### C. Heat-capacity

In order to further address the question of the ground state magnetic ordering, the C behavior as a function of temperature in zero magnetic field as well as in the presence of magnetic fields are shown in Fig. 4a. There is a distinct upturn below 19 K with a prominent peak at 15 K, confirming bulk magnetic ordering. C however does not vary as $T^3$, a functional form expected for antiferromagnetics (without any gap effects) well below 15 K, but varies rather linearly below 8 K. Possibly, this is an artifact of the formation of magnetic gap (see below). Focusing now on the H-dependent C(T) curves, the peak shifts to lower temperatures as H is applied, which endorses that the magnetic ordering below 15 K is in fact antiferromagnetic. All the curves intersect at a characteristic temperature, about 17 K [5,6]. From these data, the information about magneto-caloric effect behavior has been obtained by deriving (Fig. 4b and 4c) the entropy change, $\Delta S$= S(H)-S(0), and adiabatic change in temperature ($\Delta T$= T(S,H)-T(S,0)). The noteworthy finding is that, in addition to the positive (negative) peak in $\Delta S$ ($\Delta T$) around 12 K known for antiferromagnets, there is a peak with the opposite sign at the onset of magnetic ordering, around 17 K. This is consistent [5] with the fact that, in a very narrow temperature interval above 15 K, ferromagnetic correlations are dominant, endorsing the inferences from $\chi(T)$ data above.

### D. Electrical resistance and magnetoresistance

The temperature coefficient of ρ is found to be positive for the range 20-300 K with rather small ρ values, thereby establishing that this compound is a metal. Figure 5a shows the data for the low temperature range only, as the curve above 50 K is otherwise featureless. Below 17 K, there is a very weak upturn of ρ (in zero H) resulting in a minimum in ρ(T) followed by a kink at 15 K and a smooth fall thereafter. The upturn below 17K can be explained by proposing that there is a formation of magnetic Brillouin-zone boundary gap, which implies that the magnetic structure (in a narrow temperature range) proposed above is of a modulated-type as in some rare-earth metals [7] and this gap effect apparently gets reduced. A small applied field shifts the minimum to a lower temperature (e.g., to 13 K for H= 5 kOe) and thus the onset of magnetic order is very sensitive to H. The features due to magnetic ordering are smoothened for larger applications of H (say, 10 and 30 kOe). The magnetoresistance, defined as MR= [$\rho(H)$- $\rho(0)$]/ $\rho(0)$, is negative and the magnitude is reasonably large for small applications of H, as shown in figure 5b and the negative sign of MR with an observable magnitude persists over a wide temperature range in the paramagnetic state, as noted for few other exceptional heavy rare-earth compounds [8]. Another noteworthy finding is that there is an upward curvature in the plot of MR(H) for H>30 kOe at low temperatures (see figure 5b), the origin of which is not clear. Possibly, the magnetic gaps undergo complex changes with H.

## IV. CONCLUSION

The formation of a ternary Eu compound $EuCu_2As_2$ and its magnetic behavior are reported. While the compound appears to undergo antiferromagnetic ordering below 15 K, the magnetic behavior in a narrow temperature interval before the onset of antiferromagnetism seems ferromagnetic-like as measured by magnetic susceptibility. In this connection, it may be recalled that, in many compounds with layered structure [for example, see Refs. 9 and 10], the intra-layer interaction has been established to be ferromagnetic, while the interlayer interaction is antiferromagnetic. Therefore, it is possible that the dominant, nearest-neighbour (that is, intra-layer) Eu-Eu interaction is ferromagnetic as indicated by a positive value of $\theta_p$ the magnitude of which is also comparable to that of magnetic ordering temperature. However, the net magnetic structure could be antiferromagnetic due to inter-layer antiferromagnetic action; if this interlayer coupling is comparatively very weak, then, at the onset of magnetic ordering, a small magnetic field (of the magnitude employed in the present $\chi$ studies) may be enough to disrupt this arrangement to induce three-dimensional ferromagnetism. It is of interest to carry out careful neutron diffraction experiments on this compound to address this aspect better. In any case, the results reveal that there is a delicate balance between ferromagnetism and antiferromagnetism in this compound and a theoretical investigation is called for to address such a close balance.

**Acknowledgement:** We thank Kartik K Iyer for his help during measurements.

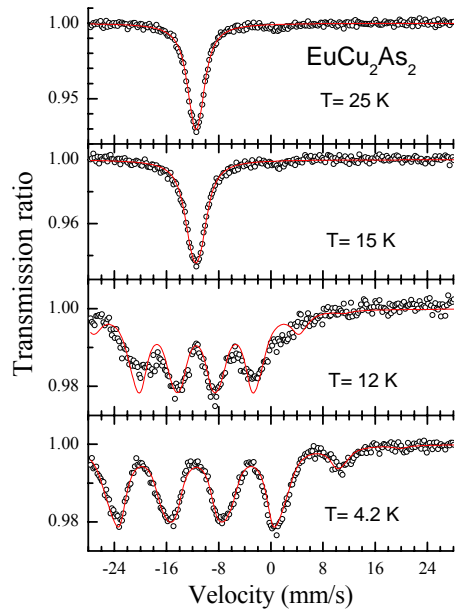

Fig. 1: (color online) $^{151}$Eu Mössbauer spectra of EuCu$_2$As$_2$ at low temperatures. The continuous lines represent least-square fit of the data to a single Lorentzian line at 25 K and to a hyperfine-split pattern at 12 and 4.2 K.

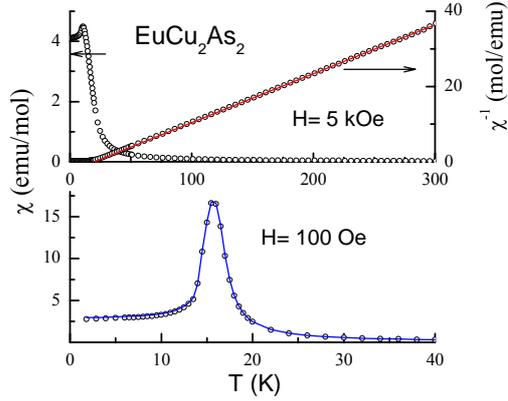

Fig. 2: (Color online) Magnetic susceptibility obtained in a field of 5 kOe and 100 Oe for $EuCu_2As_2$. The plot of inverse $\chi(T)$ (H= 5 kOe) is also shown and the continuous line represents linear region. The continuous line through the data points in the case H= 100 Oe represents the behavior for the field-cooled (FC) condition of the specimen; otherwise all the data points correspond to the zero-field-cooled state of the specimen.

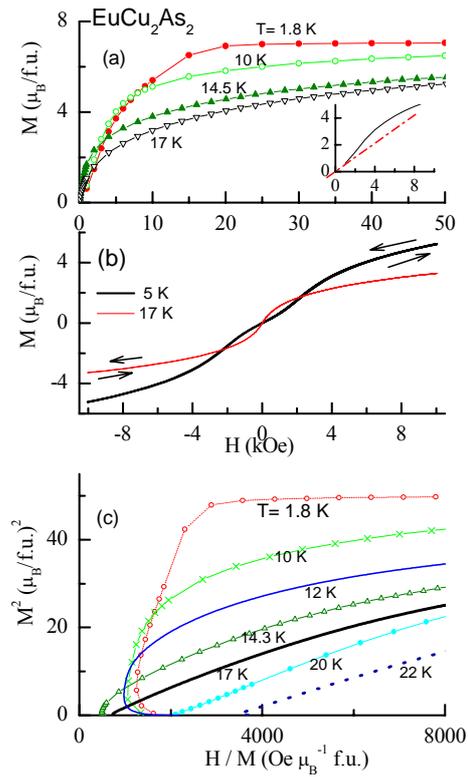

Fig. 3: (color online) (a) Isothermal magnetization behavior $EuCu_2As_2$ at 1.8, 10, 14.5 and 17 K. The lines through the data points serve as a guide to the eyes. The inset shows the low-field data in an expanded region to highlight metamagnetic transition and the dashed line is obtained by linear extrapolation of the data near zero-field region, (b) Hysteresis loops at 5 and 17 K, and (c) Arrott plot, $M^2$ vs H/M, for $EuCu_2As_2$.

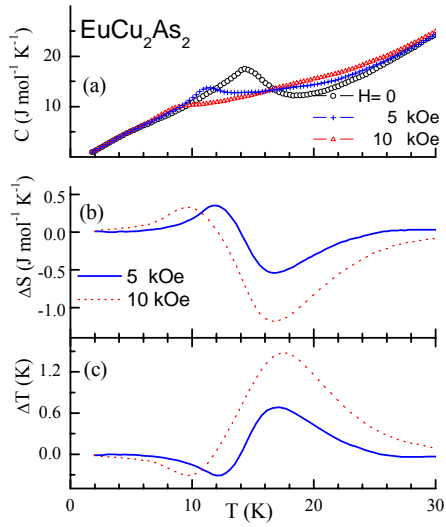

Fig. 4: (color online) (a) Heat capacity, (b) entropy change, and (c) adiabatic change in temperature, as a function of temperature for $EuCu_2As_2$.

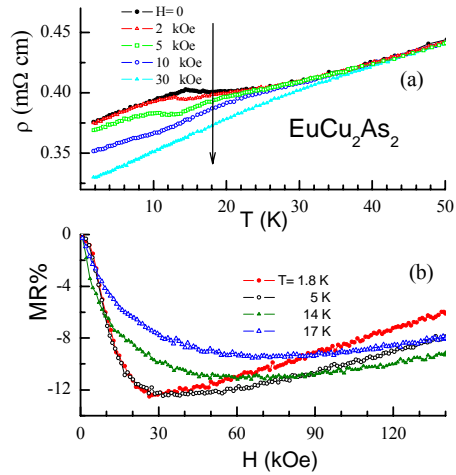

Fig. 5: (color online) (a) Electrical resistivity ($\rho$) as a function of temperature (1.8 – 50 K) in the presence of various externally applied magnetic fields and (b) magnetoresistance, defined as $[\rho(H)- \rho(0)]/ \rho(0)$, as a function of external field at selected temperatures for $EuCu_2As_2$.